\documentclass[11pt,a4paper]{article}
\usepackage{cite}

\usepackage{amsmath, amsthm,mathtools,url,amssymb,upgreek,slashed,dsfont}

\usepackage{mathabx} %\widebar
\renewcommand{\bar}{\widebar}

\usepackage{ifpdf}
\ifpdf
  \usepackage[pdftex]{graphicx}
  \usepackage{epstopdf}
\else
  \usepackage[dvips]{graphicx}
\fi
\parskip=\baselineskip

\numberwithin{equation}{section}

\newcommand{\beq}{\begin{equation}}
\newcommand{\eeq}{\end{equation}}

\newcommand{\CC}{\mathbb{C}}
\newcommand{\RR}{\mathbb{R}}

\newcommand{\SL}{\mathrm{SL}}
\newcommand{\SO}{\mathrm{SO}}

\renewcommand{\sl}{\mathfrak{sl}}
\newcommand{\so}{\mathfrak{so}}

\renewcommand{\gg}{\mathfrak{g}}
\newcommand{\hh}{\mathfrak{h}}
\newcommand{\kk}{\mathfrak{k}}

\DeclareMathOperator{\tr}{Tr}

\newcommand{\id}{\mathds{1}}

\newcommand{\chibar}{ {\bar{\chi}} }
\newcommand{\psibar}{ {\bar{\psi}} }

\renewcommand{\d}{\mathrm{d}}
\newcommand{\DD}{\mathcal{D}}
\newcommand{\dd}{\mathrm{D}}

\newcommand{\Dslash}{\slashed{D}}

\newcommand{\ddslash}{\slashed{\dd}}

\renewcommand{\S}{\mathrm{S}}

\newcommand{\e}{\varepsilon}

\newcommand{\ee}{\epsilon}
\newcommand{\eebar}{\bar{\epsilon}}

\newcommand{\ts}{\thinspace}
\newcommand{\nts}{\negthinspace}

\newcommand{\Y}{\mathrm{Y}}

\renewcommand{\phi}{\varphi}

\renewcommand{\t}{\zeta}

\usepackage[symbol]{footmisc}

\renewcommand{\thefootnote}{\fnsymbol{footnote}}

\begin{document}

\begin{titlepage}

\begin{flushright}
\end{flushright}

\vskip 1.5in

\begin{center}
{\bf\Large{A Lax Operator for $d=2$ $N=2$ Supergravity}}

\vskip 0.5cm {Robert F. Penna\footnote[1]{rpenna@ias.edu} } 
\vskip 0.05in {\small{ \textit{Institute for Advanced Study}
\vskip -.4cm
{\textit{Einstein Drive, Princeton, NJ 08540 USA}}}
}
\end{center}
\vskip 0.5in
\baselineskip 16pt

\begin{abstract}  

General relativity and supergravity become integrable systems after dimensional reduction to two spacetime dimensions.  This means the equations of motion can be encoded in the flatness condition for a Lax operator.  Nicolai and Warner found Lax operators for dimensionally reduced supergravity.  They gave explicit formulas primarily for the case with $N=16$ supersymmetry in two dimensions (which corresponds to $N=8$ supergravity in four dimensions).  In this note, we derive analogous results for the case with $N=2$ supersymmetry in two dimensions (which corresponds to $N=1$ supergravity in four dimensions).   This is the simplest example of the general fact that supergravity becomes an integrable system after dimensional reduction to two  dimensions.  

\end{abstract}

%\date{\today}

\end{titlepage}

\renewcommand*{\thefootnote}{\arabic{footnote}}
\setcounter{footnote}{0}

\setcounter{tocdepth}{1}
\tableofcontents

\section{Introduction}
\label{sec:intro}

Lax operators are one of the basic objects in the theory of integrable systems.  A Lax operator is a Lie algebra valued 1-form, $L$, obeying the flatness  condition
\beq\label{eq:flat}
dL + L \wedge L = 0 \,.
\eeq
The flatness condition encodes the equations of motion of an integrable system. $L$ is a function of the spacetime coordinates and an auxiliary parameter, $\t \in \CC$, called the spectral parameter.  

The Lax operator can also be formulated as a $(0,1)$ gauge field, $A$, on twistor space \cite{woodhouse1988geroch,mason1996integrability}.  
The flatness condition \eqref{eq:flat} becomes the partial flatness condition
\beq\label{eq:pflat}
\bar{\partial} A + A \wedge A = 0 \,.
\eeq
A conceptual advantage of this formulation is that now the spacetime coordinates and the spectral parameter are all just coordinates on twistor space, which is six real dimensional.  An interesting recent development is the observation \cite{costellotalk,bittleston2020twistors,penna2020twistor} that \eqref{eq:pflat} is the equation of motion of a Chern-Simons theory on twistor space.  Thus the problem of finding and classifying integrable systems is closely related to the problem of finding and classifying solutions of a Chern-Simons theory on twistor space.

General relativity and supergravity become integrable systems after dimensional reduction to two spacetime dimensions.  In the case of pure general relativity, the dimensionally reduced theory is an $\SL(2,\RR)/\SO(2)$ coset sigma model coupled to a dilaton and 2d general relativity  \cite{
geroch1972method,
breitenlohner1987geroch,
nicolai1991two,
schwarz1995classical,schwarz1995classicalb,
lu2007infinite,lu2007infiniteb}.  The 2d theory has an infinite dimensional Kac-Moody symmetry called the Geroch group.  (The Poisson brackets of the charges generate a Yangian symmetry \cite{korotkin1998yangian,koepsell1999yangian}.)  Under the action of the Geroch group, the Minkowski metric can be mapped to any 4d vacuum metric with two commuting Killing vectors.   In principle, this reduces the problem of deriving vacuum metrics with two commuting Killing vectors to pure algebra.   For a recent example of these facts in action, see \cite{penna2021einstein}, where we gave an algebraic derivation of the Einstein-Rosen metric, which describes a cylindrical gravitational wave.

In the case of supergravity, the dimensionally reduced theory is a coset sigma model coupled to a dilaton, matter fermions, and 2d supergravity.   Nicolai \cite{nicolai1987integrability,nicolai1994new} and Nicolai and Warner \cite{nicolai1989structure} found Lax operators for these models.  They described the Lax operator explicitly for the case with $N=16$ supersymmetry in two dimensions (which corresponds  to $N=8$ supersymmetry in four dimensions).  In the present paper, we will derive the Lax operator for the case with $N=2$ supersymmetry in two dimensions\footnote{The $N=2$ Lax operator described herein misses some fermion terms in the equations of motion that are not missed by the Lax operators with $N>2$.  This can be traced to the fact that $\SO(2)$ is abelian.  This issue was first noted by Nicolai \cite{nicolai1991two}.  We comment on this issue at the end of Section \ref{sec:lax}.}  (which corresponds to $N=1$ supersymmetry in four dimensions).

The 2d metric is
\beq\label{eq:2dmetric}
ds^2 = \lambda^2 ( -dt^2 + dr^2 ) \,,
\eeq
where $\lambda = \lambda(t,r)$.  We eliminated the off diagonal components and set $-g_{tt} = g_{rr} = \lambda^2$ using a diffeomorphism.   It turns out that the equation of motion for $\lambda$ decouples from the other equations of motion.  So we can  consider the Lax operator with or without $\lambda$.   The Lax operators that have been derived so far do not include $\lambda$ and we will not include it here either.  Extending the Lax operator to include the equation of motion for $\lambda$ is an interesting open problem\footnote{
In the case of pure general relativity, there is a formula for $\lambda$ in terms of a cocycle on the Geroch group \cite{breitenlohner1987geroch}.}.  Nicolai \cite{nicolai1994new} has extended the Lax operator to include the off diagonal components of \eqref{eq:2dmetric}.  For simplicity, we will not consider that extension in this work.  

In a similar way, it is possible to decouple the equations of motion for the 2d gravitinos.  
Nicolai \cite{nicolai1991two}  described an $N=2$ Lax operator that does not include the gravitinos.   
The $N=16$ Lax operators of \cite{nicolai1989structure,nicolai1994new} do include the gravitinos.  
We will include the gravitinos in our Lax operator\footnote{
The dimensional reduction of $N=1$ supergravity from four dimensions to a single null dimension has a hyperbolic Kac-Moody symmetry \cite{nicolai1992hyperbolic} that mixes the matter fermions and the gravitinos.  This is one motivation for keeping the gravitinos in the  Lax operator. 
}.

There is a remarkable interplay between supersymmetry and integrability in these models (first observed in \cite{nicolai1987integrability}).  The action and the equations of motion have quadratic and quartic fermion terms.  But the Lax operator only has quadratic fermion terms.  The quartic fermion terms in the equations of motion come from the $L\wedge L$ term in the flatness condition \eqref{eq:flat}.  
The upshot is that we can (and do) ignore quartic fermion terms throughout this work, and yet still obtain the Lax operator of the full theory (the theory with quartic fermion terms).  It would be good to have a better conceptual understanding of why this works.  It would also be interesting to find a similar principle at the level of the action, relating the quartic fermion terms in the action to the quadratic fermion terms, perhaps using twistor Chern-Simons theory  \cite{costellotalk,bittleston2020twistors,penna2020twistor}.

This paper is relatively self contained.   We begin in three spacetime dimensions.  We check that the 3d action is supersymmetric to linear order in the fermions. Then we derive the 2d action and equations of motion by dimensionally reducing the 3d action.  We discuss the Lax operator  \eqref{eq:L1}--\eqref{eq:L3}  in Section \ref{sec:lax}.  Our conventions for the metric, curvature tensors, gamma matrices, etc. follow the textbook \cite{freedman2012supergravity}.

\section{$N=2$ Supergravity in Three Dimensions}
\label{sec:3daction}

We begin by introducing $N=2$ supergravity in three spacetime dimensions \cite{marcus1983three,marcus1984infinite,de1993locally}.   This action can be derived by dimensionally reducing $N=1$ supergravity from four dimensions.  However, the reduction to three dimensions requires some nontrivial redefinitions of the 4d fermions (for a similar example, see \cite{de1986d}), so it is easier to begin in three dimensions.

The pure supergravity part of the action is
\beq\label{eq:Sgrav}
\frac{1}{2} \int \d^3x \ts e R
 -\frac{1}{2} \int \d^3x \ts e \ts \psibar^I_m \gamma^{mnp} D_n \psi^I_p  \,.
\eeq
The first term is the Einstein-Hilbert term and the second term is the gravitino kinetic term.   The metric signature is $(-++)$.   
The fields are the frame field, $e_m^a$, and a pair of gravitinos, $\psi_m^I$, $I=1,2$. The gravitinos are Majorana fermions.  In the first term, $e=\det e_m^a$ and $R$ is the Ricci scalar.  
 In the second term, the covariant derivative is
\beq\label{eq:covD}
D_m \psi_n^I = \partial_m \psi_n^I + \tfrac{1}{4}\omega_{mab}\gamma^{ab}\psi_n^I + \dots
\eeq
The dots indicate an additional coupling to be defined below (equation \ref{eq:Dpsi}).  $\omega_{mab}$ is the spin connection.  We use $m,n,p,\dots$ for curved indices and $a,b,c,\dots$ for flat indices.  

The matter part of the action includes an $\SL(2,\RR)/\SO(2)$ coset sigma model.   To set this up, let $U\in \SL(2,\RR)$ be the basic field of the sigma model.  The Lie algebra of $\SL(2,\RR)$, $\gg=\sl(2,\RR)$, decomposes as 
\beq
\gg = \hh \oplus \kk \,,
\eeq
where $\hh = \so(2)$ and $\kk$ is the orthogonal complement of $\so(2)$ in $\sl(2,\RR)$.
Define Lie algebra valued fields, $Q_m \in \hh$ and $P_m \in \kk$, by
\beq\label{eq:QPdef}
-(\partial_m U) U^{-1} = Q_m + P_m \,.
\eeq
$Q_m$ transforms as an $\SO(2)$ gauge field under local $\SO(2)$ transformations, $\delta U = - hU$, $h\in \so(2)$.  
To see why, note that the variation of \eqref{eq:QPdef} is
\begin{align}
-(\partial_m hU) U^{-1}  + (\partial_m U) U^{-1} h
	&= - \partial_m h  - h (\partial_m U) U^{-1} +  (\partial_m U) U^{-1} h \notag\\
	&= - \partial_m h - [P_m , h ] \,.
\end{align}
Collect terms according to the Lie algebra decomposition $\gg=\hh\oplus\kk$ to get
\begin{align}
\delta Q_m &= -(\partial_m h) \,, \label{eq:dQ}\\
\delta P_m &= - [P_m,h] \,.
\end{align}
The first line is the transformation law for an $\SO(2)$ gauge field.

Now the matter part of the action is
\beq\label{eq:Smatter}
-\frac{1}{2} \int \d^3x \ts e \ts g^{mn} \tr (P_m P_n) 
-\frac{1}{2} \int \d^3x \ts e \ts \chibar^I \Dslash \chi^I \,.
\eeq
The first term is the sigma model kinetic term and the second term is the matter fermion kinetic term.  In the first term, $g^{mn} = e^m_a e^n_b \eta_{ab}$ is the inverse metric ($\eta_{ab}$ is the 3d Minkowski metric).  The matter fermions, $\chi^I$, $I=1,2$, are Majorana fermions.  The slash operator is $\Dslash \chi^I = \gamma^m D_m \chi^I$. 
The covariant derivative, $D_m \chi^I$, is 
\beq
D_m \chi^I	= \partial_m \chi^I + \tfrac{1}{4}\omega_{mab}\gamma^{ab} \chi^I + \tfrac{3}{2}Q_m^3 \e^{IJ}\chi^J \,. \label{eq:Dchi}
\eeq
$\e^{IJ}$ is the Levi-Civita symbol ($\e^{12} = -\e^{21} = 1$).  The factor of $Q_m^3$ in \eqref{eq:Dchi} is defined by $Q_m = Q_m^3 \Y^3$, where $\Y^3$ is the generator of $\so(2)$ (see equation \ref{eq:Ys}).  In what follows, we will usually suppress the Lie algebra index on $Q_m^3$ for brevity.  
The covariant derivative of the gravitino \eqref{eq:covD} is defined similarly:
\beq
D_m \psi_n^I	= \partial_m \psi_n^I + \tfrac{1}{4}\omega_{mab}\gamma^{ab}\psi_n^I + \tfrac{1}{2}Q_m \e^{IJ}\psi^J_n \,.  \label{eq:Dpsi}
\eeq

The sum of the gravity and matter actions, \eqref{eq:Sgrav} and \eqref{eq:Smatter}, is not supersymmetric.  In fact, it is not even supersymmetric to linear order in the fermions.  To get  an action that is supersymmetric to linear order in the fermions, we need to add one more term.  First, define the $\sl(2,\RR)$ basis
\beq
\Y^1 = \begin{pmatrix} 1	&	0 \\	0	&	-1 \end{pmatrix} , \quad
\Y^2 = \begin{pmatrix} 0	&	1 \\	1	&	0 \end{pmatrix} , \quad
\Y^3 = \begin{pmatrix} 0	&	1 \\	-1	&	0 \end{pmatrix} . \label{eq:Ys}
\eeq
In this basis, the components of $P_n$ are
\beq
P_n =  P_n^1 \Y^1 + P_n^2 \Y^2 = P_n^A \Y^A \quad (A=1,2) \,.
\eeq
Let $t^A_{IJ} = \frac{1}{\sqrt{2}} \Y^A_{IJ}$ and $t_{IJ} = t^A_{IJ} \Y^A$ ($A=1,2$).  The additional term we need is
\beq
- \int \d^3x \ts e \ts t^A_{IJ} \nts \left( \chibar^I \gamma^m \gamma^n \psi^J_m \right) \nts P^A_n \,.
\eeq
In the next section, we will check that the action 
\beq\label{eq:Stot}
S = S_0 + S_{1/2} + S_{3/2} + S_2 + S_N \,,
\eeq
with
\begin{align}
S_0		&= -\frac{1}{2} \int \d^3x \ts e \ts g^{mn} \tr (P_m P_n) \,, \label{eq:S0}\\
S_{1/2}	&= -\frac{1}{2} \int \d^3x \ts e \ts \chibar^I \Dslash \chi^I  \,, \label{eq:Schi}\\
S_{3/2}	&= -\frac{1}{2} \int \d^3x \ts e \ts \psibar^I_m \gamma^{mnp} D_n \psi^I_p \,, \label{eq:Spsi}\\
S_2		&= \frac{1}{2} \int \d^3x \ts e R \,, \label{eq:S2}\\
S_N		&=  - \int \d^3x \ts e \ts t^A_{IJ} \nts \left( \chibar^I \gamma^m \gamma^n \psi^J_m \right) \nts P^A_n \,, \label{eq:SN}
\end{align}
is supersymmetric to linear order in the fermions.

First, we record some facts about $Q_m$ and $P_m$ that will be needed in what follows.  Define 
\begin{align}
D_m U		&= \partial_m U + Q_m U \,, \label{eq:DU}\\
D_m U^{-1}	&= \partial_m U^{-1} - U^{-1} Q_m \,, \label{eq:DUinverse}
\end{align}
and
\begin{align}
D_m Q_n	&= \partial_m Q_n \,, \label{eq:DQ}\\
D_m P_n	&= \partial_m P_n + [Q_m, P_n] \,. \label{eq:DP}
\end{align}
There is no $[Q_m,Q_n]$ term in \eqref{eq:DQ} because $Q_m, Q_n \in \so(2)$.  Equations \eqref{eq:DU}--\eqref{eq:DP} are compatible with definition \eqref{eq:QPdef} because
\begin{align}
-D_m \nts \left( (\partial_n U) U^{-1} \right)
	& = -\partial_m \left( (\partial_n U) U^{-1} \right)
		-Q_m (\partial_n U) U^{-1} + (\partial_n U) U^{-1} Q_m \notag \\
	&= -\partial_m \left((\partial_n U) U^{-1} \right) - \left[Q_m , \partial_n U U^{-1}\right] \notag \\
	&= \partial_m Q_n + \partial_m P_n + [Q_m, P_n] \notag \\
	&= D_m Q_n + D_m P_n \,.
\end{align}

We are going to need the important pair of equations
\begin{align}
\partial_m Q_n  -  \partial_n Q_m + [P_m , P_n] &= 0 \,, \label{eq:Qeq} \\
D_m P_n - D_n P_m  &= 0 \,. \label{eq:Peq}
\end{align}
To derive these equations, first combine \eqref{eq:DU} and \eqref{eq:QPdef} to get
\beq
(D_m + P_m) U = \partial_m U + Q_m U + P_m U = 0  \,.
\eeq
Then expand the commutator $[D_m+P_m, D_n + P_n]U$ to get
\begin{align}
[D_m+P_m, D_n + P_n] U		&= [D_m, D_n] U + [D_m, P_n] U + [P_m, D_n] U + [P_m,P_u] U  \notag \\
						&= (\partial_m Q_n) U - (\partial_n Q_m) U 
							+(D_m P_n) U - (D_n P_m) U  + [P_m,P_n] U \notag \\
						&= 0 \,.
\end{align}
Collect the terms on the middle line according to the Lie algebra decomposition $\gg=\hh\oplus \kk$ to get  \eqref{eq:Qeq}--\eqref{eq:Peq}.

Finally, define
\beq
\DD_m \psi_n^I = \partial_m \psi_n^I + \tfrac{1}{4}\omega_{mab}\gamma^{ab}\psi_n^I \,.
\eeq
$D_m$ and $\DD_m$ are related by 
\beq
D_m \psi_n^I	= \DD_m \psi_n^I +  \tfrac{1}{2}Q_m \e^{IJ}\psi^J_n \,.
\eeq
An important identity is
\beq
[D_m,D_n]\psi_p^I = [\DD_m,\DD_n]\psi_p^I - \e^{AB} \e^{IJ}  P_m^A P_n^B \psi^J_p \,. \label{eq:DD}
\eeq
To prove this, first expand
\begin{align}
[D_m,D_n] \psi_p^I	=& D_m D_n \psi_p^I - D_n D_m \psi_p^I \notag\\
				=& D_m (\DD_n \psi_p^I + \frac{1}{2} Q_n \e^{IJ} \psi_p^J)
					-D_n (\DD_m \psi_p^I + \frac{1}{2}Q_m \e^{IJ} \psi_p^J) \notag\\
				=& \DD_m \DD_n \psi_p^I + \frac{1}{2} Q_m \e^{IJ} \DD_n \psi_p^J
					+\frac{1}{2}(D_m Q_n) \e^{IJ}\psi^J_p + \frac{1}{2}Q_n \e^{IJ}D_m \psi_p^J \notag\\
				&- \DD_n \DD_m \psi_p^I - \frac{1}{2} Q_n \e^{IJ} \DD_m \psi_p^J
					-\frac{1}{2}(D_n Q_m) \e^{IJ} \psi^J_p -\frac{1}{2}Q_m \e^{IJ} D_n \psi^J_p \notag \\
				=& [\DD_m,\DD_n]\psi_p^I + \frac{1}{2}(D_mQ_n - D_n Q_m) \e^{IJ} \psi_p^J \,.
\end{align}
Then use $D_mQ_n = \partial_m Q_n$ (equation \ref{eq:DQ}) to get
\beq
[D_m,D_n] \psi_p^I	=  [\DD_m,\DD_n]\psi_p^I + \frac{1}{2}(\partial_m Q_n - \partial_n Q_m) \e^{IJ} \psi_p^J \,. \label{eq:DD1}
\eeq
Now recall the remark beneath \eqref{eq:Dchi}.  The term in parentheses is 
\beq
\partial_m Q_n^3 - \partial_n Q_m^3	=  - [P_m,P_n]^3 = -2 P_m^A P_n^B \e^{AB} \,.  \label{eq:DD2}
\eeq
The first equality follows from equation \eqref{eq:Qeq}.  The second equality follows upon writing $P_m = P_m^A \Y^A$ and using $[\Y^A,\Y^B] = 2\e^{AB} \Y^3$.
Plugging \eqref{eq:DD2} into \eqref{eq:DD1} gives \eqref{eq:DD}

\section{Supersymmetry}
\label{sec:susy} 

In this section, we will check that the 3d action \eqref{eq:Stot}--\eqref{eq:SN} is supersymmetric to linear order in the fermions.  
It is not invariant to cubic order in the fermions.  To get complete invariance, we would need to add quartic fermion terms to the action.

The supersymmetry variations of the gravity fields are
\begin{align}
\delta e^a_m 		&= \frac{1}{2} \eebar^I \gamma^a \psi^I_m \,, \label{eq:deltae}\\
\delta \psi^I_m 		&= D_m \ee^I \,. \label{eq:deltapsi}
\end{align}
The infinitesimal parameters, $\ee^I$, $I=1,2$, are Majorana fermions.  
The covariant derivative of an infinitesimal paramter is
\beq
D_m \ee^I	= \partial_m \ee^I + \tfrac{1}{4}\omega_{mab}\gamma^{ab}\ee^I + \tfrac{1}{2}Q_m \e^{IJ}\ee^J \,.
\eeq
The variations of the inverse tetrad and determinant are
\beq
\delta e^m_a = -\frac{1}{2}\eebar^I \gamma^m \psi^I_a \,, \quad
\delta e = \frac{1}{2} \ts e \ts \eebar^I \gamma^p \psi^I_p \,. \label{eq:deltae2}
\eeq

The variations of the matter fields are
\begin{align}
\delta \chi^I 		&= -\gamma^m \ee^J P_m^A \ts t^A_{IJ} \,, \label{eq:deltachi}\\
(\delta U U^{-1})^A 	&= \frac{1}{2} t^A_{IJ} \eebar^I \chi^J \,.  \label{eq:deltaU}
\end{align}
The second line is the variation of $U$, the basic field of the sigma model.  
The variations of $Q_m$ and $P_m$ are obtained by varying \eqref{eq:QPdef},
\begin{align}
\delta Q_m + \delta P_m	&= -(\partial_m \delta U ) U^{-1} + (\partial_m U) U^{-1} \delta U U^{-1} \notag \\
					&= - \partial_m \left(\delta U U^{-1}\right) 
						+ \left[(\partial_m U) U^{-1},\delta U U^{-1}\right] \notag \\
					&= - \partial_m \left(\delta U U^{-1}\right) 
						- \left[Q_m , \delta U U^{-1}\right]
						- \left[P_m , \delta U U^{-1}\right] .  \label{eq:dQP}
\end{align}
Collecting terms according to the Lie algebra decomposition $\sl(2,\RR) = \so(2) \oplus \kk$ gives
\begin{align}
\delta Q_m &= - \left[P_m , \delta U U^{-1} \right]  \,, \label{eq:deltaQ} \\
\delta P_m &= - \partial_m \left(\delta U U^{-1}\right)  - \left[Q_m , \delta U U^{-1}\right] 
	\equiv - D_m \negthinspace \left(\delta U U^{-1} \right) . \label{eq:deltaP}
\end{align}
These equations and \eqref{eq:deltaU}  define the variations of $Q_m$ and $P_m$.

Our goal is to show that the 3d action \eqref{eq:Stot}--\eqref{eq:SN} is invariant under \eqref{eq:deltae}--\eqref{eq:deltapsi} and \eqref{eq:deltachi}--\eqref{eq:deltaU} to linear order in the fermions.  
There are two parts to this problem.
In the first part, we will show that the terms in the variation that are linear in $\chi^I$ sum to zero.   
These terms come from varying $U$ in $S_0$, varying $\chi^I$ in $S_{1/2}$, and varying $\psi^I$ in $S_N$.
In the second part, we will show that the terms in the variation that are linear in $\psi_m^I$ sum to zero.  These terms come from varying $e_m^a$ in $S_0$ and $S_2$, varying $\psi^I_m$ in $S_{3/2}$, and varying $\chi^I$ in $S_N$.
Table \ref{tab:susy} summarizes all the terms that appear in the variation of the action.
 
\begin{table}[]
\begin{center}
\begin{tabular}{ |c|ccccc| } 
\hline
				& $\delta S_0$	& $\delta S_{1/2}$	& $\delta S_{3/2}$	& $\delta S_2$	& $\delta S_N$	\\ 
\hline
$\delta e^a_m$		& 1			& 3				& 3				& 1			& 3		\\
$\delta U$			& 1			& 3				& 3				& ---			& 3		\\ 
$\delta \chi^I$		& ---			& 1				& ---				& ---			& 1		\\
$\delta \psi^I_m$	& ---			& --- 				& 1				& ---			& 1		\\
\hline
\end{tabular}
\caption{Varying $S_i$ gives terms that are linear (1) and cubic (3) in the fermions.}
\label{tab:susy}
\end{center}
\end{table}

Varying $U$ in $S_0$ gives
\begin{align}
-\int \d^3 x \ts e \ts g^{mn} \tr(P_m \delta P_n )
	&= -2 \int \d^3 x \ts e \ts g^{mn} P_m^A \delta P_n^A \notag \\
	&=  \int \d^3 x \ts e \ts g^{mn} P_m^A D_n \nts \left(t^A_{IJ} \eebar^I \chi^J\right) \notag \\
	&= -\int \d^3 x \ts e \ts g^{mn}\nts \left(t^A_{IJ} \eebar^I \chi^J\right)\nts D_n P_m^A  \notag \\
	&= -\int \d^3 x \ts e \ts \gamma^m \gamma^n\nts \left(t^A_{IJ} \eebar^I \chi^J\right)\nts D_n P_m^A \,.\label{eq:S0U}
\end{align}
In the first step, we used $\tr (\Y^A \Y^B) = 2 \delta^{AB}$.
In the second step, we used equation \eqref{eq:deltaP} for $\delta P_n^A$.
In the third step, we integrated by parts. 
In the last step, we used $g^{mn} = \gamma^m \gamma^n - \gamma^{mn}$ and $\gamma^{mn} D_m P_n = 0$ (which follows from equation \ref{eq:Peq}).

Varying $\chi^I$ in $S_{1/2}$ gives
\begin{align}
-\int \d^3 x \ts e \left(\delta \chibar^I\right) \Dslash \chi^I
	&= -\int \d^3 x \ts e \left(t_{IJ}^A \eebar^J \gamma^m P_m^A \right) \Dslash \chi^I \notag\\
	&= \int \d^3 x \ts e \left( t_{IJ}^A  \left(D_n \eebar^J\right) \gamma^m P^A_m \right) \gamma^n \chi^I
		+ \int \d^3 x \ts e \left(t_{IJ}^A  \eebar^J \gamma^m \left( D_n P_m^A \right) \right) \gamma^n \chi^I \notag \\
	&= \int \d^3 x \ts e \ts t^A_{IJ}  \left( \chibar^I \gamma^n \gamma^m \left(D_n \ee^J  \right) \right)\nts P^A_m 
		+ \int \d^3 x \ts e \left(t^A_{IJ} \eebar^J \gamma^m \gamma^n  \chi^I\right)\nts D_n P_m^A \notag \\
	&= \int \d^3 x \ts e \ts t^A_{IJ} \left( \chibar^I \gamma^m \gamma^n \left( D_m \ee^J \right) \right) \nts P_n^A 
		 + \int \d^3 x \ts e \ts \gamma^m \gamma^n \left(t^A_{IJ} \eebar^I \chi^J\right)\nts D_n P_m^A \,. \label{eq:Schichi}
\end{align}
In the second step, we integrated by parts.  
In the last step, we used $\gamma^m \gamma^n = g^{mn} + \gamma^{mn}$ and $\gamma^{mn} D_m P_n = 0$ to pull $\gamma^m \gamma^n$ outside the fermion bilinear in the second term.

Varying $\psi_m^J$ in $S_N$ gives
\beq\label{eq:SNpsi}
-\int \d^3 x \ts e \ts t^A_{IJ} \left( \chibar^I \gamma^m \gamma^n (D_m \ee^J)  \right) P_n^A \,.
\eeq
Now it is clear that equations \eqref{eq:S0U}--\eqref{eq:SNpsi} sum to zero.  We thus confirm that the terms in the variation of the action that are linear in $\chi^I$ sum to zero.

Now consider the terms in the variation of the action that are linear in $\psi_m^I$.  These terms come from varying $e_m^a$ in $S_0$ and $S_2$, varying $\psi^I_m$ in $S_{3/2}$, and varying $\chi^I$ in $S_N$.

Varying $e_m^a$ in $S_0$ gives
\begin{align}
-\frac{1}{2} \int \d^3 x & \left[ \left( \frac{1}{2} e \ts \eebar^I \gamma^p \psi_p^I \right) g^{mn}
	+ 2 e \left( -\frac{1}{2}\eebar^I \gamma^m \psi_a^I \right) e^{na} \right] \tr(P_m P_n) \notag \\
	&= -\frac{1}{4} \int \d^3 x \ts e \ts \eebar^I \left( \gamma^p \psi_p^I g^{mn}-2\gamma^m \psi^{In}\right)\tr(P_m P_n) \notag \\
	&= -\frac{1}{2} \int \d^3 x \ts e \ts \eebar^I  \left( \gamma^p \psi_p^I g^{mn}-2\gamma^m \psi^{In}\right) P_m^A P_n^A \,. \label{eq:susy2a}
\end{align}
The first equality is straightforward, we are just collecting some common factors in the integrand.  The second equality follows upon expanding $P_m = P_m^A \Y^A$ and using $\tr\left(\Y^A \Y^B\right) = 2 \delta^{AB}$.

Varying $\psi^I_m$ in $S_{3/2}$ and integrating by parts gives
\begin{align}
-\frac{1}{2}  \int \d^3x \ts e \left(D_m \eebar^I \right) &\gamma^{mnp} D_n \psi^I_p 
	 -\frac{1}{2} \int \d^3x  \ts e \ts \psibar_m^I \gamma^{mnp} D_n \left(D_p \ee^I \right) \notag \\
	&= \frac{1}{2} \int \d^3 x \ts e \ts  \eebar^I \gamma^{mnp} [D_m,D_n] \psi_p^I \,. \label{eq:psivar}
\end{align}
The commutator is \eqref{eq:DD}
\beq
[D_m,D_n]\psi_p^I = [\DD_m,\DD_n]\psi_p^I - \e^{AB} \e^{IJ}  P_m^A P_n^B \psi^J_p \,.
\eeq
It is a standard result that the contribution from $[\DD_m,\DD_n]\psi_p^I$ is cancelled by the variation of $e_m^a$ in $S_2$.  We are interested in the contribution from the second term,
\beq
 -\frac{1}{2} \int \d^3 x \ts e \ts \e^{AB} \e^{IJ} \nts \left( \eebar^I \gamma^{mnp} \psi_p^J \right) \nts P_m^A P_n^B  
	= \frac{1}{2} \int \d^3 x \ts e \ts \e^{AB} \e^{IJ} \nts \left (\eebar^I \gamma^m \gamma^p \gamma^n \psi_p^J \right)\nts P_m^A P_n^B  \,. \label{eq:susy2b}
\eeq
We have used an identity,
 $\gamma^{mnp} \e^{AB} P_m^A P_n^B = -\gamma^m\gamma^p\gamma^n \e^{AB}  P_m^A P_n^B$,
whose proof is straightforward but somewhat tedious (see Appendix \ref{sec:app}).

Varying $\chi^I$ in $S_N$ gives
\beq\label{eq:deltachiSN}
-\int \d^3 x \ts e \ts t^A_{IJ} \nts \left( \left(\eebar^K \gamma^p P_p^B t_{IK}^B \right) \gamma^m \gamma^n \psi_m^J \right)\nts P_n^A 
 = -\int \d^3 x \ts e \ts t^A_{IJ} t_{IK}^B \nts \left( \eebar^K \gamma^p  \gamma^m \gamma^n \psi_m^J \right)\nts P_p^B P_n^A \,.
\eeq
Expand $t^A_{IJ} t^B_{IK}$ using 
\beq\label{eq:ttidentity}
t^A_{IJ} t^B_{IK} = \frac{1}{2} \delta^{AB} \delta_{JK} + \frac{1}{2}\e^{AB}\e_{JK} \,.
\eeq
Equation \eqref{eq:ttidentity} is easy to check by a direct calculation (there are 16 components to check).  Plugging \eqref{eq:ttidentity} into \eqref{eq:deltachiSN}
gives two terms.  The first term is
\begin{align}
-\frac{1}{2}\int \d^3 x\ts e \left( \eebar^I \gamma^p \gamma^m \gamma^n \psi_m^I \right)\nts P_p^A P_n^A 
&=-\frac{1}{2}\int \d^3 x\ts e \left( \eebar^I \gamma^m \gamma^p \gamma^n \psi_p^I \right)\nts P_m^A P_n^A \notag \\
&= \frac{1}{2} \int \d^3 x \ts e \ts \eebar^I \nts \left(\gamma^p \psi_p^I g^{mn} - 2\gamma^m \psi^{In}\right) \nts P_m^A P_n^A \,.\label{eq:susy2c}
\end{align}
The first equality is just an index relabeling.  The second equality follows from the elementary identity
$\gamma^m \gamma^p \gamma^n =  - \gamma^p g^{mn}  + 2 \gamma^n g^{mp}  - \gamma^p \gamma^{mn}$.

The contribution from the second term of \eqref{eq:ttidentity} is
\beq
- \frac{1}{2} \int \d^3 x \ts e \ts \e^{AB} \e^{IJ} \nts \left( \eebar^J \gamma^m \gamma^p \gamma^n \psi^I_p \right) \nts P_m^B P_n^A
=	- \frac{1}{2}\int \d^3 x \ts e \ts \e^{AB} \e^{IJ} \nts \left( \eebar^I \gamma^m \gamma^p \gamma^n \psi^J_p \right) \nts P_m^A P_n^B \,. \label{eq:susy2d}
\eeq
This is also just an index relabeling.

Now it is clear that equations \eqref{eq:susy2a}, \eqref{eq:susy2b}, \eqref{eq:susy2c}, and \eqref{eq:susy2d} sum to zero.  We have therefore shown that the 3d action \eqref{eq:Stot}--\eqref{eq:SN} is invariant under \eqref{eq:deltae}--\eqref{eq:deltapsi} and \eqref{eq:deltachi}--\eqref{eq:deltaU} to linear order in the fermions.

\section{Dimensional Reduction}
\label{sec:dimreduce}

In this section, we will derive the action of $N=2$ supergravity in two spacetime dimensions by dimensionally reducing the 3d action \eqref{eq:Stot}--\eqref{eq:SN}.  

The 3d metric is
\beq\label{eq:metric}
ds^2 = \lambda^2 (-dt^2 + dr^2 ) + \rho^2 (dx^2)^2 \,,
\eeq
where $\lambda=\lambda(t,r)$ and $\rho=\rho(t,r)$ are functions of $t$ and $r$ only.  The off diagonal components have been eliminated, and the diagonal components $ - g_{tt} = g_{rr} = \lambda^2$ have been identified, using the equations of motion and a diffeomorphism.  We dimensionally reduce along $x^2$.

\subsection{$S_0$}

The first term in the 3d action is the sigma model kinetic term,
\beq
S_0 =  -\frac{1}{2} \int \d^3x \ts e \ts g^{mn} \tr (P_m P_n)  \,.
\eeq
Assume the fields are independent of $x^2$.  Then $P_2 = 0$ (equation \ref{eq:QPdef}).   So $S_0$ becomes
\beq\label{eq:S0reduce}
-\frac{1}{2} \int \d^3x \ts e \ts g^{\mu\nu} \tr (P_\mu P_\nu)  \,,
\eeq
where $\mu,\nu=t,r$ are 2d indices.
The frame field is $e^a_m = {\rm diag}(\lambda,\lambda,\rho)$ and its determinant is $e=\lambda^2 \rho$.   
The metric and inverse metric are
\beq\label{eq:3dmetric}
g_{mn} = 
\begin{pmatrix}
\lambda^{2} h_{\mu\nu}	&	0	\\
0					&	\rho^{2} 
\end{pmatrix} , \quad
g^{mn} = 
\begin{pmatrix}
\lambda^{-2} h^{\mu\nu}	&	0	\\
0					&	\rho^{-2} 
\end{pmatrix} ,
\eeq
where $h_{\mu\nu} = {\rm diag}(-1,1)$ is the 2d Minkowski metric.  So the 2d sigma model kinetic term is
\beq
\S_0 =  -\frac{1}{2} \int \d^2x \rho\ts h^{\mu\nu} \tr (P_\mu P_\nu) \,.
\eeq

\subsection{$S_{1/2}$}

The next term in the 3d action is the matter fermion kinetic term,
\beq\label{eq:Shalf}
S_{1/2}	= -\frac{1}{2} \int \d^3x \ts e \ts \chibar^I \gamma^m D_m \chi^I  \,.
\eeq
The covariant derivative is \eqref{eq:Dchi}
\beq\label{eq:Dchi2}
D_m \chi^I	= \partial_m \chi^I 
				+ \tfrac{1}{4}\omega_{mab} \gamma^{ab} \chi^I + \tfrac{3}{2}Q_m  \e^{IJ} \chi^J \,.
\eeq
The spin connection can be eliminated in two spacetime dimensions.
To see why, we need to compute the spin connection of the 3d metric \eqref{eq:metric}.

The first step is to compute the anholonomy coefficients,
\beq
\Omega_{[mn]p} = (\partial_m e_n^a - \partial_n e_m^a)e_{ap} \,,
\eeq
where $e_{am} = {\rm diag}(-\lambda,\lambda,\rho)$.
The nonvanishing components are
\beq
\Omega_{[tr]t} = \lambda \partial_r \lambda \,, \quad 
\Omega_{[tr]r} = \lambda \partial_t \lambda  \,, \quad 
\Omega_{[t2]2} = \rho \partial_t \rho \,, \quad
\Omega_{[r2]2} = \rho \partial_r \rho \,.
\eeq

The next step is to compute the spin connection,
\beq
\omega_{m[np]} = \frac{1}{2}\left(\Omega_{[mn]p} - \Omega_{[np]m} + \Omega_{[pm]n}\right) .
\eeq
The nonvanishing components of $\omega_{mab}$ are
\beq\label{eq:omegacmpts}
\omega_{t[tr]} = -\lambda^{-1} \partial_r \lambda \,, \quad
\omega_{r[tr]} = -\lambda^{-1} \partial_t \lambda  \,, \quad 
\omega_{2[t2]} = -\lambda^{-1} \partial_t \rho \,, \quad
\omega_{2[r2]} = -\lambda^{-1} \partial_r \rho \,.
\eeq
We stress that these are the components of $\omega_{mab}$ with one curved index and two flat indices.  We could use dots to distinguish curved and flat indices (for example, $\omega_{\dot{t}tr}$ versus $\omega_{\dot{t}\dot{t}\dot{r}}$), but the extra notation becomes cumbersome.  Instead, we will simply state whether indices are curved or flat when confusion might arise.

The spin connection term in equation \eqref{eq:Dchi2} is
\begin{align}\label{eq:wab}
\frac{1}{4} \omega_{mab} \gamma^m  \gamma^{ab} 
	&= \frac{1}{4} e^m_c \omega_{mab} \gamma^c  \gamma^{ab} \notag \\
	&= \frac{1}{2} \lambda^{-1}  \left(\omega_{ttr} \gamma^t \gamma^{tr} + \omega_{rtr}  \gamma^r  \gamma^{tr} \right)
		+\frac{1}{2}\rho^{-1} \left( \omega_{2 t 2} \gamma^2 \gamma^{t2} 
		+ \omega_{2 r 2} \gamma^2  \gamma^{r2} \right) .
\end{align}
To simplify the gamma matrices that appear here (and in what follows), it is convenient to introduce a matrix representation.
First recall the Pauli matrices,
\beq
\sigma_1 = \begin{pmatrix} 0 & 1 \\ 1 & 0 \end{pmatrix} , \quad
\sigma_2 = \begin{pmatrix} 0 & -i \\ i & 0 \end{pmatrix} , \quad
\sigma_3 = \begin{pmatrix} 1 & 0 \\ 0 & -1 \end{pmatrix} .
\eeq
The (anti)commutators of the Pauli matrices are $\{ \sigma_i , \sigma_j \} = 2\delta_{ij} \id$ and $[\sigma_i,\sigma_j] = 2i \e_{ijk} \sigma_k$.  
The matrix representation of the gamma matrices we will use is
\begin{align}\label{eq:gammarep}
\gamma^t = i \sigma_1 \,, \quad \gamma^r = \sigma_2 \,, \quad \gamma^2 = \sigma_3 \,, \\
\gamma_t = -i \sigma_1 \,, \quad \gamma_r = \sigma_2 \,, \quad \gamma_2 = \sigma_3 \,.
\end{align}
Now it is straightforward to compute
\beq\label{eq:gammaids1}
\gamma^t \gamma^{tr} = -\gamma^r \,, \quad 
\gamma^r \gamma^{tr} = -\gamma^t \,, \quad
\gamma^2 \gamma^{t2} = -\gamma^t \,, \quad 
\gamma^2 \gamma^{r2} = -\gamma^r \,.
\eeq
Plugging into \eqref{eq:wab} gives
\beq\label{eq:omegasimple}
\frac{1}{4} \omega_{mab} \gamma^m  \gamma^{ab} 
	= \frac{1}{2} (\lambda^{-2}\gamma^\mu \partial_\mu \lambda +  \rho^{-1} \lambda^{-1} \gamma^\mu \partial_\mu \rho ) \,.
\eeq
(Recall that $\mu=t,r$ is a 2d index.)  This is the contribution of the spin connection to the derivative \eqref{eq:Dchi2}.

Now consider the contribution of the spin connection to the integrand  \eqref{eq:Shalf},
\begin{align}\label{eq:Shalf2}
e \ts \chibar^I \gamma^m D_m \chi^I 
	&= e \chibar^I \nts \left( e^m_c \gamma^c \partial_m \right) \nts \chi^I
		  + e \ts \chibar^I \nts \left(\tfrac{1}{4} \omega_{mab} \gamma^m \gamma^{ab} \right) \nts \chi^I
		  + e \chibar^I \nts \left(\tfrac{3}{2}Q_m  \e^{IJ} \gamma^m\right) \nts \chi^J \notag \\
	&= (\lambda^{1/2} \rho^{1/2} \chibar^I) \gamma^\mu \partial_\mu (\lambda^{1/2} \rho^{1/2} \chi^I) 
		+ (\lambda^{1/2} \rho^{1/2} \chibar^I) (\tfrac{3}{2}Q_\mu  \e^{IJ} \gamma^\mu) (\lambda^{1/2} \rho^{1/2} \chi^J ) \notag \\
	&\equiv  (\lambda^{1/2} \rho^{1/2} \chibar^I) \gamma^\mu \dd_\mu (\lambda^{1/2} \rho^{1/2} \chi^I) \,.
\end{align}
In the last step, we defined 
\beq 
\dd_\mu (\lambda^{1/2} \rho^{1/2} \chi^I) =
	\partial_\mu (\lambda^{1/2} \rho^{1/2} \chi^I)
	+\tfrac{3}{2}Q_\mu  \e^{IJ} (\lambda^{1/2} \rho^{1/2} \chi^J ) \,.
\eeq
The spin connection has now been eliminated.  The key step was moving $\lambda^{1/2} \rho^{1/2}$ inside the derivative in equation \eqref{eq:Shalf2}.

To eliminate $\lambda$ from the matter fermion kinetic term, rescale $\lambda^{1/2} \chi^I \rightarrow \chi^I$.  This gives the 2d matter fermion kinetic term,
\beq
\S_{1/2} = -\frac{1}{2} \int \d^2 x \rho^{1/2} \chibar^I \ddslash \big(\rho^{1/2} \chi^I \big) \,. 
\eeq

\subsection{$S_{3/2}$}

The next term in the 3d action is the gravitino kinetic term \eqref{eq:Spsi}
\beq\label{eq:Spsi2}
S_{3/2}	= -\frac{1}{2} \int \d^3x \ts e \ts \psibar^I_m \gamma^{mnp} D_n \psi^I_p \,.
\eeq
This is the most complicated term.  Fortunately, most of the steps are similar to steps from the previous two subsections.  The first step is to replace the 3d indices $m,n,p,\dots$ with 2d indices $\mu,\nu,\rho,\dots$
The gravitino becomes (compare \ref{eq:3dmetric})
\beq
\psi_m^I = (\lambda \psi_\mu^I, \rho \psi_2^I) \,.
\eeq
The rank 3 Clifford algebra element becomes
\beq
\gamma^{[mn2]} = e^{-1} \gamma^{[\mu\nu2]} \,.
\eeq
And the gravitino action becomes
\beq\label{eq:psipsi}
S_{3/2} = -\frac{1}{2} \int \d^3x \left[ 
		\lambda \psibar_\mu^I \gamma^{\mu 2 \nu} D_2 (\lambda \psi_\nu^I)
		+\lambda \psibar_\mu^I \gamma^{\mu\nu 2}D_\nu (\rho \psi_2^I) 
		+ \rho \psibar_2^I \gamma^{2\mu\nu} D_\mu (\lambda \psi_\nu^I) \right] .
\eeq

It turns out \cite{nicolai1987integrability,nicolai1989structure,nicolai1994new}  we can use the equations of motion to set
(compare \ref{eq:metric})
\beq\label{eq:psipsi1}
\psi_\mu^I = \gamma_\mu \psi^I \,.
\eeq
Plugging into \eqref{eq:psipsi}, integrating by parts, and using $\gamma_\mu \gamma^{\mu2\nu} = \gamma^{2\nu\mu} \gamma_\mu=  \gamma^{2\nu}$, gives
\beq\label{eq:psipsi3}
S_{3/2} = \frac{1}{2} \int \d^3x \left[
	\lambda \psibar^I \gamma^{2 \nu} D_2 (\lambda \gamma_\nu \psi^I)
	+ 2\lambda \psibar^I \gamma^{\nu 2}D_\nu (\rho \psi_2^I) \right].
\eeq
The first term on the rhs vanishes.  To see why, note that $Q_2 = 0$ because the fields are independent of $x^2$ (equation \ref{eq:Qeq}).  So the only possible contribution to this term is from the spin connection.
But
\begin{align}
\gamma^{2\nu} \omega_{2ab} \gamma^{ab} \gamma_\nu 
	& = -2 \omega_{2t2} \gamma^\nu \gamma^2 \gamma^{t2} \gamma_\nu 
	-2 \omega_{2r2} \gamma^\nu \gamma^2 \gamma^{r2} \gamma_\nu \notag \\
	& = 2  \omega_{2t2} \gamma^\nu \gamma^t \gamma_\nu 
	+ 2 \omega_{2r2} \gamma^\nu \gamma^r \gamma_\nu \notag  \\
	& = 0 \,.
\end{align}
The second equality follows from \eqref{eq:gammaids1} and the third equality follows from 
$\gamma^\nu \gamma^\mu \gamma_\nu = 0$.
So the gravitino kinetic term is
\beq\label{eq:psipsi4}
S_{3/2} 
	= \int \d^3x \thinspace  \lambda \psibar^I \gamma^{\mu 2}D_\mu (\rho \psi_2^I) 
	= -\int \d^3x \thinspace  \lambda \psibar^I \gamma^2 \gamma^\mu D_\mu (\rho \psi_2^I) \,.
\eeq

At this point, $S_{3/2}$ is essentially no more complicated than the matter fermion kinetic term of the previous subsection.  As in that case, the contribution from the spin connection can be eliminated.
The covariant derivative in \eqref{eq:psipsi4} is
\beq
D_\mu(\rho \psi_2^I) = 
	\partial_\mu(\rho \psi_2^I) 
	+ \frac{1}{4}\omega_{\mu a b} \gamma^{ab} \rho \psi_2^I 
	+ \frac{1}{2}Q_\mu \e^{IJ} \rho \psi_2^J \,.
\eeq
The spin connection contributes (compare \ref{eq:omegasimple})
\beq
\frac{1}{4}\omega_{\mu a b} \gamma^\mu \gamma^{ab}  
	= \frac{1}{2}\lambda^{-1} \gamma^\mu \partial_\mu \lambda \,.
\eeq
So the integrand of $S_{3/2}$ is
\begin{align}
-\lambda \psibar^I \gamma^2 \gamma^\mu D_\mu (\rho \psi_2^I) 
	&=	- (\lambda^{1/2} \psibar^I ) \gamma^2 \gamma^\mu \partial_\mu (\lambda^{1/2} \rho \psi_2^I)
		- \frac{1}{2} \e^{IJ}  \left(\lambda^{1/2} \psibar^I \right) \gamma^2 \gamma^\mu Q_\mu  
			\left(\lambda^{1/2} \rho \psi_2^J\right) \notag \\
	&\equiv - \lambda^{1/2} \psibar^I \gamma^2 \gamma^\mu \dd_\mu (\lambda^{1/2} \rho \psi_2^I) \,.
\end{align}
In the last step, we defined
\beq
\dd_\mu (\lambda^{1/2} \rho \psi_2^I) 
	=  \partial_\mu (\lambda^{1/2} \rho \psi_2^I) + \frac{1}{2}   \e^{IJ} Q_\mu (\lambda^{1/2} \rho \psi_2^J) \,.
\eeq
Rescaling $\lambda^{1/2} \psi^I \rightarrow \psi^I$ and $\lambda^{1/2} \psi_2^I \rightarrow \psi_2^I$ gives
\beq
\S_{3/2} = - \int \d^2 x \ts \psibar^I \gamma^2 \ddslash (\rho \psi_2^I) \,.
\eeq
This is the 2d gravitino kinetic term.

\subsection{$S_2$}

The 3d Einstein-Hilbert term is
\beq\label{eq:S22}
S_2	 = \frac{1}{2} \int \d^3x \ts e R \,.
\eeq
We need to compute the scalar curvature, $R$, of the 3d metric \eqref{eq:metric}.
We have already computed the spin connection.
The next step is to compute the Riemann curvature tensor,
\beq
R_{mnab} = \partial_m \omega_{nab} - \partial_n \omega_{mab} 
	+ \omega_{mac} {{\omega_n}^c}_b - \omega_{nac} {{\omega_m}^c}_b \,.
\eeq
The nonvanishing independent components are
\begin{align}
R_{trtr}	&= -\partial_t (\lambda^{-1} \partial_t \lambda) + \partial_r (\lambda^{-1} \partial_r \lambda ) \,, \\
R_{t2t2}	&= -\partial_t (\lambda^{-1} \partial_t \rho) + (\lambda^{-1} \partial_r \lambda )(\lambda^{-1} \partial_r \rho ) \,, \\
R_{t2r2}	&= -\partial_t (\lambda^{-1} \partial_r \rho) + (\lambda^{-1} \partial_r \lambda )(\lambda^{-1} \partial_t \rho ) \,, \\
R_{r2r2}	&= -\partial_r (\lambda^{-1} \partial_r \rho) + (\lambda^{-1} \partial_t \lambda )(\lambda^{-1} \partial_t \rho ) \,.
\end{align}
These are the components of $R_{mnab}$ with two curved indices and two flat indices.

The Ricci curvature is $R_{mn} = {{R_m}^p}_{np}$.  
So the integrand of the Einstein-Hilbert term is
\begin{align}\label{eq:S2terms12}
e R	&= e (g^{tt} R_{tt} + g^{rr} R_{rr} + g^{22}R_{22})  \notag \\
	&= \lambda^2 \rho (-\lambda^{-2} R_{tt} + \lambda^{-2} R_{rr} + \rho^{-2} R_{22}) \notag \\
	&= -2 h^{\mu\nu} \rho \partial_\mu (\lambda^{-1} \partial_\nu \lambda) 
		- 2h^{\mu \nu} \partial_\mu \partial_\nu \rho \,.
\end{align}
The second term does not contribute because it is a total derivative.  So the 2d Einstein-Hilbert term is
\beq
S_2 = -\int \d^2 x \rho h^{\mu\nu} \partial_\mu (\lambda^{-1} \partial_\nu \lambda) \,.
\eeq

\subsection{$S_N$}

The last term in the 3d action is the interaction term \eqref{eq:SN},
\beq
S_N =  - \int \d^3x \ts e \ts t^A_{IJ} \nts \left( \chibar^I \gamma^m \gamma^n \psi^J_m \right) \nts P^A_n \,.
\eeq
$P_2^A = 0$ because the fields are independent of $x^2$.  So 
\beq
\gamma^m \gamma^n \psi_m^J P_n^A 
	= \frac{1}{\lambda} \gamma^\mu \gamma^\nu \gamma_\mu \psi^J P_\nu^A + \frac{1}{\lambda} \gamma^2 \gamma^\nu \psi_2^J P_\nu^A \,.
\eeq
The first term vanishes because 
$\gamma^\mu \gamma^\nu \gamma_\mu \psi^J = 0$.
This leaves
\beq
S_N = - \int \d^3x \ts \rho \lambda \ts t^A_{IJ} \nts \left( \chibar^I \gamma^2 \gamma^\nu \psi^J_2 \right) \nts P^A_\nu \,.
\eeq
Rescaling $\lambda^{1/2} \chi^I \rightarrow \chi^I$ and $\lambda^{1/2} \psi_2^I \rightarrow \psi_2^I$ gives
\beq
\S_N = -\int \d^2 x \ts \rho \ts t^A_{IJ} \left( \chibar^I \gamma^2 \gamma^\nu \psi_2^J \right) \nts P_\nu^A \,.
\eeq
This is the 2d interaction term.

\section{$N=2$ Supergravity in Two Dimensions}
\label{sec:2daction}

In the previous section, we obtained the action for $N=2$ supergravity in two spacetime dimensions.  Now we will derive the equations of motion.  The 2d action is
\beq
\S = \S_0 + \S_{1/2} + \S_{3/2} + \S_2 + \S_N \,,
\eeq
with
\begin{align}
\S_0		&= -\frac{1}{2} \int \d^2x \ts \rho h^{\mu\nu} \tr(P_\mu  P_\nu) \,, \\
\S_{1/2} 	&= -\frac{1}{2} \int \d^2 x \ts \rho^{1/2} \chibar^I \ddslash \big(\rho^{1/2} \chi^I \big) \,, \\
\S_{3/2} 	&= - \int \d^2 x \ts \psibar^I \gamma^2 \ddslash (\rho \psi_2^I) \,, \\
\S_2		&= - \int \d^2 x \ts \rho h^{\mu\nu} \partial_\mu (\lambda^{-1} \partial_\nu \lambda) \,, \\
\S_N		&= - \int \d^2 x \ts \rho \ts t^A_{IJ} \left( \chibar^I \gamma^2 \gamma^\mu \psi_2^J \right) \nts P_\mu^A \,.
\end{align}
The derivative,  $\ddslash = \gamma^\mu \dd_\mu$, is an $\SO(2)$ covariant derivative, with
\begin{align}
\dd_\mu \chi^I		&= \partial_\mu \chi^I + \tfrac{3}{2} Q_\mu \e^{IJ} \chi^J \,, \\
\dd_\mu \psi^I		&= \partial_\mu \psi^I + \tfrac{1}{2} Q_\mu \e^{IJ} \psi^J \,, \\
\dd_\mu \psi_2^I	&= \partial_\mu \psi_2^I + \tfrac{1}{2} Q_\mu \e^{IJ} \psi_2^J \,.
\end{align}
The spin connection was eliminated in the previous section.

\subsection{Fermion Equations of Motion}

It is convenient to integrate by parts and use $\gamma^\mu \gamma^2 = -\gamma^2 \gamma^\mu$ to get
\beq
\S_{3/2} 	= - \int \d^2 x \ts \psibar^I \gamma^2 \ddslash (\rho \psi_2^I)  
		= -\int \d^2 x \ts \rho \psibar_2^I \gamma^2 \left(\ddslash \psi^I\right) .
\eeq
A Majorana flip and some index relabeling gives
\beq
\S_N		= - \int \d^2 x \ts \rho \ts t^A_{IJ} \left( \chibar^I \gamma^2 \gamma^\mu \psi_2^J \right) \nts P_\mu^A 
		= \int \d^2 x \ts \rho \ts t^A_{IJ} \left( \psibar_2^I \gamma^2 \gamma^\mu \chi^J \right) P_\mu^A \,.
\eeq
Now it is clear that varying the action with respect to $\psibar_2^I$ gives the equation of motion
\beq\label{eq:eompsi}
\ddslash \psi^I = t^A_{IJ} \gamma^\mu \chi^J P_\mu^A \,.
\eeq

Varying the action with respect to $\psibar^I$ gives the equation of motion
\beq\label{eq:eompsi2}
\ddslash (\rho \psi_2^I) = 0 \,.
\eeq
And varying the action with respect to $\chibar^I$ gives the equation of motion
\beq\label{eq:eomchi}
\rho^{-1/2} \ddslash (\rho^{1/2} \chi^I) = - t^A_{IJ} \gamma^2 \gamma^\mu \psi_2^J P_\mu^A  \,.
\eeq
Equations \eqref{eq:eompsi}--\eqref{eq:eomchi} are the equations of motion for the fermions\footnote{The 3d action gives an additional equation of motion for $\psi_2$ that is not visible in the 2d action in this gauge \eqref{eq:psipsi1}.  Nicolai \cite{nicolai1994new} has shown how to include the missing equation of motion in the $N=16$ Lax operator.  We will ignore this extension for simplicity.}.

\subsection{Scalar Equations of Motion}

Varying the action with respect to $\lambda$  gives
\beq\label{eq:eomrho}
h^{\mu\nu} \partial_\mu \partial_\nu \rho = 0 \,.
\eeq
Varying the action with respect to $\rho$ gives
\beq
-h^{\mu\nu} \tr(P_\mu P_\nu) 
	- \frac{1}{2} \rho^{-1/2} \chibar^I \ddslash (\rho^{1/2} \chi^I)
	- \psibar_2^I \gamma^2 \left( \ddslash \psi^I \right)
	- h^{\mu\nu} \partial_\mu (\lambda^{-1} \partial_\nu \lambda)
	+ t^A_{IJ} \psibar_2^I \gamma^2 \gamma^\mu \chi^J P_\mu^A = 0 \,.
\eeq
The third and fifth terms cancel (equation \ref{eq:eompsi}).  The factor of $\rho^{1/2}$ in the second term can be pulled outside the derivative because $\chibar^I \gamma^\mu \chi^I=0$.  So the equation of motion for $\lambda$ becomes
\beq\label{eq:eomlambda}
h^{\mu\nu} \partial_\mu (\lambda^{-1} \partial_\nu \lambda) 
 	= -h^{\mu\nu}\tr(P_\mu P_\nu) - \frac{1}{2} \chibar^I \Dslash \chi^I \,.
\eeq
The equation of motion obtained by varying $h_{\mu\nu}$ is equivalent to \eqref{eq:eomlambda}.
 
The last and most complicated equation of motion is the equation of motion for $P_\mu$.  It is obtained by varying the action with respect to $U$, the basic field of the sigma model (see equation \ref{eq:QPdef}).  The variations of $Q_\mu$ and $P_\mu$ are  related to $\delta U$ by equations \eqref{eq:deltaQ} and \eqref{eq:deltaP},
\begin{align}
\delta Q_\mu &= - \left[P_\mu , \delta U U^{-1} \right]  \,, \\
\delta P_\mu &= - D_\mu \negthinspace \left(\delta U U^{-1} \right) \label{eq:deltaP2} \,. 
\end{align}
Writing $Q_\mu = Q_\mu^3 \Y^3$ gives
\beq\label{eq:deltaQ3}
\delta Q_\mu^3 = 2\e^{AB}   (\delta U U^{-1})^A P_\mu^B \,.
\eeq

Now consider the variation of each term in the action with respect to $U$.
The variation of $\S_0$ with respect to $U$ is
\begin{align}
-\int \d^2 x \ts \rho \ts h^{\mu\nu} \tr(P_\mu \delta P_\nu ) 
	&= \int \d^2 x \ts \rho h^{\mu\nu} \tr \nts \left(P_\mu D_\nu (\delta U U^{-1}) \right) \notag \\
	&= -\int \d^2 x \ts h^{\mu\nu} \tr \nts \left(D_\nu (\rho P_\mu) \delta U U^{-1} \right) \notag \\
	&= -2 \int \d^2 x \ts h^{\mu\nu} D_\nu (\rho P_\mu^A) (\delta U U^{-1})^A \,.\label{eq:deltaUS0}
\end{align}
In the second step, we integrated by parts.
In the last step, we used the $\sl(2,\RR)$ basis \eqref{eq:Ys} and $\tr(\Y^A \Y^B) = 2 \delta^{AB}$.  
 
Varying $\S_{1/2}$ with respect to $U$ gives
\beq
-\frac{3}{4} \int \d^2 x \ts \e^{IJ}  \rho \left( \chibar^I \gamma^{\mu}  \chi^J \right) \delta Q_\mu^3
	= -\frac{3}{2} \int \d^2 x \ts \e^{AB} \e^{IJ} \rho \left( \chibar^I \gamma^\mu \chi^J \right) (\delta U U^{-1})^A P_\mu^B \,.
\eeq
Varying $\S_{3/2}$ with respect to $U$ gives
\beq
 - \frac{1}{2} \int \d^2 x \ts \e^{IJ} \rho \left( \psibar^I \gamma^2 \gamma^\mu \psi_2^J \right) \delta Q_\mu^3 
	= - \int \d^2 x \ts \e^{AB} \e^{IJ} \rho \left( \psibar^I \gamma^2 \gamma^\mu \psi_2^J \right) \nts (\delta U U^{-1})^A P_\mu^B  \,.
\eeq
And varying $\S_N$ with respect to $U$ and integrating by parts gives
\beq\label{eq:deltaUSN}
 -\int \d^2 x \ts \rho \ts t^A_{IJ} \left( \chibar^I \gamma^2 \gamma^\mu \psi_2^J \right) \delta P_\mu^A 
 	= -\int \d^2 x \ts D_\mu(\rho t^A_{IJ} \chibar^I \gamma^2 \gamma^\mu \psi_2^J)(\delta U U^{-1})^A \,.
\eeq
Combining \eqref{eq:deltaUS0}--\eqref{eq:deltaUSN} gives the equation of motion for $P_\mu^A$,
\begin{align}\label{eq:Peom}
2 h^{\mu\nu} D_\mu(\rho P_\nu^A) 
	&= - \frac{3}{2} \e^{AB} \e^{IJ} \rho \left( \chibar^I \gamma^\mu \chi^J \right) P_\mu^B
 	- \e^{AB} \e^{IJ} \rho \left( \psibar^I \gamma^2 \gamma^\mu \psi_2^J \right) P_\mu^B \notag \\
	& \ts\quad - D_\mu \left(\rho t^A_{IJ} \chibar^I \gamma^2 \gamma^\mu \psi_2^J\right) .
\end{align}

 \subsection{Null Coordinates}
 
The equation of motion for $\rho$ is \eqref{eq:eomrho} 
 \beq
( -\partial_t^2 + \partial_r^2 ) \rho = 0 \,.
 \eeq
 We will set $\rho = r$ from now on.  This choice is appropriate for describing cylindrical gravitational waves.  
  
The Lax operator to be introduced in the next section is particularly simple in null coordinates, $x^\pm=t\pm r$.  In these coordinates, the action of $\gamma^2$ on the gamma matrices \eqref{eq:gammarep} is
\beq
\gamma^2 \gamma_\pm = \pm \gamma_\pm \,, \quad
\eeq
The equation of motion for $\psi_2^I$ \eqref{eq:eompsi2} is
\beq
\gamma^+ \dd_+ (r \psi_2^I) + \gamma^- \dd_- (r \psi_2^I) = 0 \,.
\eeq
Multiplying by $\gamma^2$ on the left gives
\beq
\gamma^+ \dd_+ (r \psi_2^I) - \gamma^- \dd_- (r \psi_2^I) = 0 \,.
\eeq
Adding and subtracting these two equations gives
\beq
\gamma^\pm \dd_\pm (r \psi_2^I) = 0 \,. \label{eq:eompsi2null}
\eeq

The equations of motion for $\chi$ and $\psi$ are
\begin{align}
r^{-1/2} \gamma^\pm \dd_\pm (r^{1/2} \chi^I)	&= \pm t^A_{IJ} \gamma^\pm \psi_2^J P_\pm^A \,, \label{eq:eomchinull}\\
\gamma^\pm \dd_\pm \psi^I 				&= t^A_{IJ} \gamma^\pm \chi^J P_\pm^A \,. \label{eq:eompsinull}
\end{align}

 \section{Lax Operator}
 \label{sec:lax}
 
 The Lax operator is a 1-form,
 \beq
 L = L_+ dx^+ + L_- dx^- \,.
\eeq
In the $\sl(2,\RR)$ basis \eqref{eq:Ys}, the components of the Lax operator are
\beq
L_\pm	= L_\pm^A \Y^A + L_\pm^3 \Y^3 \,.
\eeq
In this basis, the Lax operator of $d=2$ $N=2$ supergravity is
\begin{align}
L_+^A	&= \frac{\t-1}{\t+1} P_+^A + \frac{\t(\t-1)}{(\t+1)^3}t^A_{IJ} \psibar_2^I \gamma_+ \chi^J \,, \label{eq:L1}\\
L_-^A	&= \frac{\t+1}{\t-1} P_-^A + \frac{\t(\t+1)}{(\t-1)^3} t^A_{IJ} \psibar_2^I \gamma_- \chi^J \,, \label{eq:L2}\\
L_\pm^3	&= Q_\pm \pm \frac{3\t}{4(\t\pm1)^2} \e^{IJ} \chibar^I \gamma_\pm \chi^J
 			+ \frac{\t}{2(\t \pm 1)^2} \e^{IJ} \psibar_2^I \gamma_\pm \psi^J
			+ \frac{\t^2}{(\t \pm 1)^4}\e^{IJ} \psibar_2^I \gamma_\pm \psi_2^J \,. \label{eq:L3}
 \end{align}
The spectral parameter, $\t$, is spacetime dependent, with
\beq\label{eq:tau}
d\t = -\frac{\t}{2r} \left(\frac{\t-1}{\t+1} dx^+ - \frac{\t+1}{\t-1}dx^- \right) .
\eeq
Our goal is to show that the equations of motion imply
\beq\label{eq:Lflat}
dL + L \wedge L = 0 \,.
\eeq

Define a 0-form $F$ by
\beq
F dx^+ \wedge dx^- = dL + L\wedge L \,.
\eeq
$F$ admits a series expansion
\begin{align}
F &=  f^{(0)} \Y^3 + f^{(1)} \frac{\t^2+1}{(\t+1)(\t-1)} +  f^{(1')} \frac{\t}{(\t+1)(\t-1)}  \notag \\
	& \quad - f_+^{(2)} \frac{\t}{(\t-1)^2} \Y^3 -  f_-^{(2)} \frac{\t}{(\t+1)^2} \Y^3 
	- f_+^{(3)} \frac{\t(\t+1)}{(\t-1)^3}  - f_-^{(3)} \frac{\t(\t-1)}{(\t+1)^3} \notag \\
	& \quad - f_+^{(4)}  \frac{\t^2 }{(\t-1)^{4}} \Y^3  - f_-^{(4)} \frac{\t^2}{(\t+1)^{4}} \Y^3 \,.
\end{align}
Our task is to show that the coefficients vanish if and only if the equations of motion are satisfied. 
The first two coefficients, $f^{(0)}$ and $f^{(1)}$, correspond to the group theoretic equations \eqref{eq:Qeq} and \eqref{eq:Peq}.  The next coefficient, $f^{(1')}$, corresponds to the sigma model equation of motion \eqref{eq:Peom}.  The higher order coefficients, $f_\pm^{(2)}$, $f_\pm^{(3)}$, and $f_\pm^{(4)}$, correspond to the $\pm$ equations  of motion  \eqref{eq:eompsi2null}--\eqref{eq:eompsinull} for $\psi^I$, $\chi^I$, and $\psi_2^I$.

The coefficient of the constant term is 
\beq
f^{(0)} = \partial_+ Q_- - \partial_- Q_+ + [P_+,P_-]  \,.
\eeq
$f^{(0)} = 0$ is equivalent to equation \eqref{eq:Qeq}.

The coefficient of the first singular term is
\beq
f^{(1)} = D_+ P_- - D_- P_+ \,.
\eeq
$f^{(1)} = 0$ is equivalent to \eqref{eq:Peq}.

The coefficient of the next singular term is
\begin{align}\label{eq:order1terms}
f^{(1')} = &-h^{\mu\nu} r^{-1} D_\mu (r P_\nu) 
		+\frac{3}{4} \e^{AB} \e^{IJ} \nts \left(\chibar^I \gamma^\mu \chi^J\right) \nts P_\mu^A \Y^B
		+\frac{1}{2} \e^{AB}  \e^{IJ} \nts \left(\psibar^I \gamma^2 \gamma^\mu \psi_2^J\right) \nts P_\mu^A \Y^B \notag \\
	&- \frac{1}{4} r^{-1} t^{IJ} \nts \left(\psibar_2^I \gamma_+ \chi^J  \right)
	- \frac{1}{4} r^{-1} t^{IJ} \nts \left(\psibar_2^I \gamma_- \chi^J \right)
	+ \frac{1}{4} \e^{AB} \e^{IJ} \nts \left( \psibar_2^I  \gamma^\mu \psi_2^J \right) \nts P_\mu^A \Y^B \,.
\end{align}
 $f^{(1')} = 0$ is the equation of motion \eqref{eq:Peom} for $P_\mu$.  To see why, note that the second line sums to
$ - \frac{1}{2} t^{IJ} D_\mu\nts\left(r \chibar^I \gamma^2 \gamma^\mu \psi_2^J\right)$
when the fermion equations of motion are satisfied.  

The coefficients of the higher order poles are
\begin{align}
f_\pm^{(2)} 	&= -\frac{3}{8} \e^{IJ} r^{-1} \partial_\pm \nts \left(r \chibar^I \gamma^\pm \chi^J \right)
				\pm \frac{1}{4}\e^{IJ} r^{-1} \partial_\pm \nts \left(r \psibar_2^I \gamma^\pm \psi^J \right)
				\pm \e^{AB}  t^B_{IJ} \nts \left( \psibar_2^I \gamma^\pm \chi^J \right) \nts P_\pm^A \,, \label{eq:f2}\\
f_\pm^{(3)}	&= \pm  \frac{1}{2} t^{IJ} r^{-3/2}  \dd_\pm \nts \left(r^{3/2}  \psibar_2^I \gamma^\pm \chi^J \right)
				+ \frac{1}{4} \e^{AB} \e^{IJ} \nts \left( \psibar_2^I \gamma^\pm \psi_2^J \right) \nts P_\pm^A \Y^B \,,\label{eq:f3} \\
f_\pm^{(4)} 	&= \pm \frac{1}{2} \e^{IJ} r^{-2} \partial_\pm \nts \left(r^2 \psibar_2^I \gamma^\pm \psi_2^J \right) . \label{eq:f4}
\end{align}
The partial derivatives in the first and third lines should be $\SO(2)$ covariant derivatives ($\partial_\pm \rightarrow \dd_\pm$). 
This defect is particular to $N=2$ supergravity (it disappears for larger $N$).
It can be traced to the fact that the $\SO(2)$ factor in the sigma model coset space is abelian.
This issue was first observed by Nicolai \cite{nicolai1991two}.
Modulo this issue, the vanishing of \eqref{eq:f2}--\eqref{eq:f4} is equivalent to the equations of motion for the fermions \eqref{eq:eompsi2null}--\eqref{eq:eompsinull}.  To see why, use $\e^{IJ} t^A_{JK} = - \e^{AB} t^B_{IK}$ and equation \eqref{eq:ttidentity}.

\noindent{\it Acknowledgment}  This research was supported, in part, by the U.S. Department of Energy and the Sivian Fund at the Institute for Advanced Study.

\appendix

\section{Proof of Equation \eqref{eq:susy2b}}
\label{sec:app}

Equation \eqref{eq:susy2b} required an identity,
\beq\label{eq:gammamnp}
\gamma^{mnp} \e^{AB} P_m^A P_n^B = -\gamma^m\gamma^p\gamma^n \e^{AB}  P_m^A P_n^B \,.
\eeq
To prove this, first use $\gamma^{mnp} = \gamma^{[m}\gamma^n\gamma^{p]}$ and antisymmetry in $m$ and $n$ to get
\beq
 \gamma^{mnp}  \e^{AB} P_m^A P_n^B  
	= -\frac{1}{3}  
		(\gamma^m\gamma^p\gamma^n + \gamma^p\gamma^n\gamma^m + \gamma^n\gamma^m\gamma^p )
		\e^{AB}   P_m^A P_n^B \,.
\eeq
Then use $\gamma^{mp} = \frac{1}{2} [\gamma^m , \gamma^p]$ to move $\gamma^p$ to the middle of each term. So
\beq
 \gamma^{mnp}  \e^{AB} P_m^A P_n^B  
 	= -\frac{1}{3} 
		(\gamma^m\gamma^p\gamma^n 
			+ \gamma^n\gamma^p\gamma^m + 2 \gamma^{pn} \gamma^m
			+ \gamma^n\gamma^p\gamma^m + 2 \gamma^n \gamma^{mp})
		\e^{AB}    P_m^A P_n^B  \,.
\eeq
Finally, use antisymmetry in $m$ and $n$ and $\gamma^{mnp} = \tfrac{1}{2}\{\gamma^m,\gamma^{np}\}$ to get
\begin{align}
 \gamma^{mnp}  \e^{AB} P_m^A P_n^B  
	&= -\frac{1}{3}   
		(-\gamma^m\gamma^p\gamma^n 
			 - 2 \gamma^{np} \gamma^m
			 - 2 \gamma^m \gamma^{np})
		\e^{AB}  P_m^A P_n^B  \notag \\
	&=  \frac{1}{3}  
		(\gamma^m\gamma^p\gamma^n + 4 \gamma^{mnp})
		 \e^{AB}  P_m^A P_n^B  \,,	 
\end{align}
which is equivalent to \eqref{eq:gammamnp}.

\bibliographystyle{jhep}
\bibliography{lax}

\end{document}